\documentclass{article}
\usepackage{color}
\def\vol#1#2#3{{\bf {#1}} ({#2}) {#3}}
\def\NP{Nucl.~Phys. }
\def\PL{Phys.~Lett. }
\def\PR{Phys.~Rev. }
\def\PRP{Phys.~Rep. }
\def\PRL{Phys.~Rev.~Lett. }
\def\PTP{Prog.~Theor.~Phys. }

\def\IJMP{Int.~J.~Mod.~Phys. }

\def\no{\nonumber}

\def\2tvec#1#2{
\left(
\begin{array}{c}
#1  \\
#2  \\   
\end{array}
\right)}

\def\mat2#1#2#3#4{
\left(
\begin{array}{cc}
#1 & #2 \\
#3 & #4 \\
\end{array}
\right)
}

\def\Mat3#1#2#3#4#5#6#7#8#9{
\left(
\begin{array}{ccc}
#1 & #2 & #3 \\
#4 & #5 & #6 \\
#7 & #8 & #9 \\
\end{array}
\right)
}

\def\3tvec#1#2#3{
\left(
\begin{array}{c}
#1  \\
#2  \\   
#3  \\
\end{array}
\right)}

\def\4tvec#1#2#3#4{
\left(
\begin{array}{c}
#1  \\
#2  \\   
#3  \\
#4  \\
\end{array}
\right)}

\def\L{\left}
\def\R{\right}

\def\hbar{\hspace{1mm}\bar{}\hspace{-1mm}h}

\def\eqn#1{
\begin{eqnarray}
#1
\end{eqnarray}
}

\begin{document}
  \title{\bf PeV Scale Right Handed Neutrino Dark Matter in $S_4$ Flavor Symmetric extra U(1) model}

  \author{Yasuhiro Daikoku\footnote{E-mail: yasu\_daikoku@yahoo.co.jp}, 
\quad Hiroshi Okada\footnote{E-mail: hokada@kias.re.kr}
\\
  {\em Institute for Theoretical Physics, Kanazawa University, Kanazawa
  920-1192, Japan.}$^*$$^\ddagger$ \\
  {\em School of Physics, KIAS, Seoul 130-722, Korea}$^\dagger$}
\maketitle

\begin{abstract}
Recent observation of high energy neutrino in IceCube experiment
suggests existence of superheavy dark matter beyond PeV. 
We identify the parent particles of neutrino 
as two degenerated right handed neutrinos, assuming the dark matter is
the heaviest right handed neutrino. 
The $O(V_{cb})\sim O(10^{-2})$ flavor symmetry breaking accounts for the 
$O(10^{-4})$ mass degeneracy of right handed neutrinos
which is a sizable scale to explain the successful resonant leptogenesis at  the PeV scale.
At the same time, non-thermal production of the heaviest right handed neutrino
gives the right amount of dark matter for $T_{RH}\sim 10$PeV. 
The footprint of flavor symmetry is left in degenerated mass spectra
of extra Higgs multiplet and colored Higgs multiplet which may be testable
for LHC or future colliders.
\end{abstract}

\vspace{-11.5cm}
\begin{flushright}
KIAS-P15009
\end{flushright}
\newpage



\section{Introduction}

The standard model (SM) is a successful theory of gauge interactions;
however, there are many unsolved issues such as
how to generate Yukawa hierarchy, meaning of small neutrino mass,
how to generate baryon asymmetry and what dark matter is.  
Recently, IceCube reported their observation of high energy neutrinos 
which might be a hint to solve some of those problems \cite{icecube-exp}.
The most important problem of SM is how to stabilize
large hierarchy between the electroweak scale, $M_W\sim 10^2$GeV, and
the Planck scale, $M_P\sim 10^{18}$GeV against quantum corrections.
The elegant mechanism of stabilizing hierarchy is supersymmetry (SUSY) \cite{SUSY},
which is the main target of Large Hadron Collider (LHC).
The existence of light Higgs boson such as 125-126GeV \cite{higgs} supports the idea of SUSY.

In the minimal supersymmetric standard model (MSSM),
as the Higgs superfields $H^U$ and $H^D$ are vector-like under the SM gauge symmetry
$G_{SM}=SU(3)_c\times SU(2)_W\times U(1)_Y$, we can introduce $\mu$-term;
\eqn{
\mu H^U H^D,
}
in superpotential. The natural size of parameter $\mu$ is $O(M_P)$, however
$\mu$ must be $O(M_W)$ in order to allow  electroweak gauge symmetry to break.
This is so-called $\mu$-problem, which is solved by making Higgs superfields chiral
under a new $U(1)_X$ gauge symmetry. Such a model is achieved based on 
$E_6$-inspired extra U(1) model \cite{extra-u1}.  
The new gauge symmetry replaces the $\mu$-term by trilinear term;
\eqn{
\lambda SH^UH^D,
}
which is converted into effective $\mu$-term when singlet $S$ develops $O(1\mbox{TeV})$ 
vacuum expectation value (VEV) \cite{mu-problem}.
At the same time, the baryon and lepton number violating terms in MSSM are replaced by
single G-interactions;
\eqn{
GQQ+G^cU^cD^c+GU^cE^c+G^cQL,
}
where $G$ and $G^c$ are new colored superfields which 
must be introduced to cancel gauge anomaly.
Although these terms induce very fast proton decay, it can be suppressed
by a $S_4$ flavor symmetry \cite{s4u1}\cite{p-s4u1}.
Therefore flavor symmetry plays an important role in stabilizing protons in
supersymmetric model.

The $S_4$ flavor symmetry can also solve the hierarchy problem  of Yukawa couplings and the problem of flavor violating processes originated from the SUSY contribution simultaneously. The former can be achieved by not assigning the $S_4$-triplet, on the other hand the latter can be done by assigning the $S_4$ doublet for the  first and second generations of left handed quarks.
Therefore the hierarchies $m_u,m_c\ll m_t$ and $m_d,m_s\ll m_b$ are
generated by the same manner as $SU(2)_W$, as the mismatch of the sizes of representations 
among the fields in operator suppresses the coefficient of the operator.
We should keep in mind that the hierarchy $m(\mbox{quark}),m(\mbox{lepton})\ll M_P$ is generated  
by the discrepancy of the sizes of representations of $SU(2)_W$
between left handed fermion and right handed fermion.
This manner is also adopted for suppressing single G-interactions 
when $G$ and $G^c$ are assigned to be $S_4$-triplets.
From the sizes of the elements of Cabibbo-Kobayashi-Maskawa (CKM) matrix, 
the $S_3$ subgroup of $S_4$ should be broken by $O(10^{-2})$.

It is well known that heavy right handed neutrino (RHN) not only realizes small neutrino mass
by see-saw mechanism
but also generates baryon asymmetry of the universe through leptogenesis.
If the first and second generations of RHNs are assigned to be $S_4$-doublet,
then the mass degeneracy is solved to be achieved by $O(10^{-4})$.
Therefore successful resonant thermal-leptogenesis requires that
the mass of the lightest RHN is $O(\mbox{PeV})$, which
is in good agreement with the energy scale of high-energy neutrinos observed in IceCube.
Furthermore,  if we impose
constraint on the reheating temperature, $T_{RH}<10^7$GeV, for avoiding
gravitino over production \cite{reheating},
the right amount of dark matter density is obtained
through non-thermal production of the heaviest RHN. 
Although the flavor symmetry is broken, its remnant may be observed in the spectrum of extra particles
such as $G,G^c$ or extra Higgs.


\section{Symmetry Breaking}

\subsection{Gauge Symmetry}

\begin{table}[htbp]
\begin{center}
\begin{tabular}{|c|c|c|c|c|c|c|c|c|c|c|c||c|c|}
\hline
                   &$Q$ &$U^c$    &$E^c$&$D^c$    &$L$ &$N^c$&$H^D$&$G^c$    &$H^U$&$G$ &$S$ &$\Phi$&$\Phi^c$\\ \hline
$SU(3)_c$          &$3$ &$\bar{3}$&$1$  &$\bar{3}$&$1$ &$1$  &$1$  &$\bar{3}$&$1$  &$3$ &$1$ &$1$   &$1$     \\ \hline
$SU(2)_W$          &$2$ &$1$      &$1$  &$1$      &$2$ &$1$  &$2$  &$1$      &$2$  &$1$ &$1$ &$1$   &$1$     \\ \hline
$y=6Y$             &$1$ &$-4$     &$6$  &$2$      &$-3$&$0$  &$-3$ &$2$      &$3$  &$-2$&$0$ &$0$   &$0$     \\ \hline  
$6\sqrt{2/5}Q_\psi$&$1$ &$1$      &$1$  &$1$      &$1$ &$1$  &$-2$ &$-2$     &$-2$ &$-2$&$4$ &$-2$  &$2$     \\ \hline
$2\sqrt{6}Q_\chi$  &$-1$&$-1$     &$-1$ &$3$      &$3$ &$-5$ &$-2$ &$-2$     &$2$  &$2$ &$0$ &$10$  &$-10$   \\ \hline 
$x=2\sqrt{6}X$     &$ 1$&$ 1$     &$ 1$ &$2$      &$2$ &$0 $ &$-3$ &$-3$     &$-2$ &$-2$&$5$ &$0$   &$0$     \\ \hline 
$z=6\sqrt{2/5}Z$   &$-1$&$-1$     &$-1$ &$2$      &$2$ &$-4$ &$-1$ &$-1$     &$ 2$ &$ 2$&$-1$&$8$   &$-8$   \\ \hline 
$R$                &$-$ &$-$      &$-$  &$-$      &$-$ &$-$  &$+$  &$+$      &$+$  &$+$ &$+$ &$+$   &$+$     \\ \hline
\end{tabular}
\end{center}
\caption{$G_{32111}$ assignment of superfields. Where the $x$, $y$ and $z$ are charges of
$U(1)_X$, $U(1)_Y$ and $U(1)_Z$, and $Y$ is hypercharge. The charges of $U(1)_\psi$ and
$U(1)_\chi$ which are defined in Eq.(4) are also given.}
\end{table}

We extend the gauge symmetry from $G_{SM}$ to $G_{32111}=G_{SM}\times U(1)_X\times U(1)_Z$,
and add new superfields $N^c,S,G,G^c$ which are embedded in {\bf 27} representation of
$E_6$ with quark, lepton superfields $Q,U^c,D^c,L,E^c$ and Higgs superfields $H^U,H^D$.
Where $N^c$ is RHN, $S$ is $G_{SM}$ singlet and $G,G^c$ are
colored Higgs. The two U(1)s are linear combinations of 
$U(1)_\psi$ and $U(1)_\chi$ where 
$E_6\supset SO(10)\times U(1)_\psi\supset SU(5)\times U(1)_\chi\times U(1)_\psi$,
and their charges $X$ and $Z$ are given as follows
\eqn{
X=\frac{\sqrt{15}}{4}Q_\psi+\frac14 Q_\chi,\quad
Z=-\frac14 Q_\psi+\frac{\sqrt{15}}{4}Q_\chi. 
}
The charge assignments of the superfields are given in Table 1.
To break $U(1)_Z$, we add new vector-like superfields $\Phi,\Phi^c$ 
which are originated in ${\bf 351}'+\overline{{\bf 351}'}$ of $E_6$.
The invariant superpotential under these symmetries is given by
\eqn{
W_{32111}&=&W_0+W_S+W_G+W_\Phi , \\
W_0&=&Y^UH^UQU^c+Y^DH^DQD^c+Y^LH^DLE^c+Y^N H^ULN^c+Y^M\Phi N^cN^c , \\
W_S&=&kSGG^c+\lambda SH^UH^D , \\
W_G&=&Y^{QQ} GQQ+Y^{UD} G^cU^cD^c+Y^{UE} GU^cE^c+Y^{QL} G^cQL+Y^{DN}GD^cN^c , \\
W_\Phi&=&M_\Phi \Phi\Phi^c+\frac{1}{M_P}Y^\Phi(\Phi\Phi^c)^2,
}
where $M_P=2.4353\times 10^{18} \mbox{GeV}$ is reduced Planck scale and 
unimportant higher dimensional terms are omitted. Since the interactions $W_S$ 
drive squared mass of $S$ to be negative through renormalization group equations (RGEs),
spontaneous $U(1)_X$ symmetry breaking is realized and
$U(1)_X$ gauge boson $Z'$ acquires the mass
\eqn{
m(Z')\simeq \frac{5}{2\sqrt{3}}g_X\L<S\R> ,
}
where  $\L<H^{U,D}\R>\ll \L<S\R>$ is assumed based on the experimental constraint
for $Z'_\psi$ as follow
\eqn{
m(Z'_\psi)> \L\{
\begin{array}{cc}
2.51\mbox{TeV} & (\mbox{ATLAS}\cite{ATLASzprime}) \\
2.26\mbox{TeV} & (\mbox{CMS}\cite{CMSzprime}) \\
\end{array}
\R.
.
}
The constraint for $Z'$ mass is not far from this bound.
Note that the mass bound for $Z'$ depends on sparticle mass spectrum \cite{susyzprime}.
In this paper we assume $\L<H^{U,D}\R>/\L<S\R>\sim O(10^{-1})$.

If $M_\Phi=0$ in $W_\Phi$ and 
the origin of the potential $V(\Phi,\Phi^c)$ is unstable, 
then $\Phi,\Phi^c$ develop large VEVs 
along the D-flat direction of $\L<\Phi\R>=\L<\Phi^c\R>=V$, 
$U(1)_Z$ is broken and  $U(1)_Z$ gauge boson $Z''$ acquires the mass
\eqn{
m(Z'')=\frac83\sqrt{\frac52}g_ZV.
}

After the gauge symmetry breaking, since the R-parity symmetry defined by
\eqn{
R=\exp\L[\frac{i\pi}{20}(3x-8y+15z)\R] ,
}
remains unbroken, the lightest SUSY particle (LSP) is stable.

\subsection{Flavor symmetry}

\begin{table}[htbp]
\begin{center}
\begin{tabular}{|c|c|c|c|c|c|c|c|c|c|c|c|c|c|c|c|}
\hline
           &$Q_i$  &$Q_3$  &$U^c_1$&$U^c_2$&$U^c_3$&$D^c_1$&$D^c_2$&$D^c_3$ 
           &$L_1$  &$L_i$  &$E^c_1$&$E^c_2$&$E^c_3$&$N^c_i$&$N^c_3$\\
\hline
$S_4$      &$2$    &$1$    &$1$    &$1'$   &$1$    &$1'$   &$1$    &$1$     
           &$1$    &$2$    &$1$    &$1'$   &$1$    &$2$    &$1$     \\
\hline
$Z^V_9$    &$1$    &$0$    &$0$    &$0$    &$0$    &$1$    &$1$    &$0$     
           &$1$    &$0$    &$1$    &$0$    &$0$    &$0$    &$0$     \\
\hline
$Z^W_9$    &$0$    &$0$    &$0$    &$0$    &$0$    &$0$    &$0$    &$0$    
           &$0$    &$1$    &$0$    &$0$    &$0$    &$0$    &$0$      \\
\hline
$Z^P_9$    &$0$    &$0$    &$0$    &$0$    &$0$    &$0$    &$0$    &$1$    
           &$0$    &$0$    &$0$    &$0$    &$0$    &$0$    &$0$      \\
\hline
$Z_{18}$   &$0$    &$0$    &$4$    &$1$    &$0$    &$1$    &$0$    &$0$    
           &$0$    &$0$    &$1$    &$1$    &$0$    &$3$    &$2$      \\
\hline
$Z_2$      &$+$    &$+$    &$+$    &$+$    &$+$    &$+$    &$+$    &$+$    
           &$+$    &$+$    &$+$    &$+$    &$+$    &$+$    &$-$      \\
\hline
\end{tabular}
\begin{tabular}{|c|c|c|c|c|c|c|c|c|c|}
\hline
           &$H^U_i$&$H^U_3$&$H^D_i$&$H^D_3$&$S_1$&$S_2$&$S_3$ &$G$    &$G^c$\\
\hline
$S_4$      &$2$    &$1$    &$2$    &$1$    &$1$  &$1$  &$1$   &$3$    &$3$ \\
\hline
$Z^V_9$    &$0$    &$0$    &$0$    &$0$    &$0$  &$1$  &$0$   &$0$    &$0$ \\
\hline
$Z^W_9$    &$0$    &$0$    &$0$    &$0$    &$2$  &$0$  &$0$   &$0$    &$0$ \\
\hline
$Z^P_9$    &$0$    &$0$    &$0$    &$0$    &$0$  &$0$  &$0$   &$0$    &$0$ \\
\hline
$Z_{18}$   &$0$    &$0$    &$1$    &$0$    &$0$  &$3$  &$0$   &$0$    &$0$ \\
\hline
$Z_2$      &$+$    &$+$    &$+$    &$+$    &$-$  &$+$  &$+$   &$+$    &$+$ \\
\hline
\end{tabular}
\begin{tabular}{|c|c|c|c|c|c|c|c|c|c|c|}
\hline 
           &$\Phi_i$  &$\Phi_3$  &$\Phi^c_a$&$V_i$  &$X_W$&$W_i$&$P_W$&$X_P$&$P$ & $X$\\
\hline
$S_4$      &$2$       &$1$       &$3$       &$2$    &$1$  &$2$  &$1'$ &$1$  &$1'$&$1$ \\
\hline
$Z^V_9$    &$0$       &$0$       &$0$       &$-1$   &$0$  &$0$  &$0$  &$0$  &$0$ &$0$ \\
\hline
$Z^W_9$    &$0$       &$0$       &$0$       &$0$    &$-1$ &$-1$ &$-1$ &$0$  &$0$ &$0$ \\
\hline
$Z^P_9$    &$0$       &$0$       &$0$       &$0$    &$0$  &$0$  &$0$  &$-1$ &$-1$&$0$ \\
\hline
$Z_{18}$   &$1$       &$0$       &$0$       &$0$    &$0$  &$0$  &$0$  &$0$  &$0$ &$-1$\\
\hline
$Z_2$      &$+$       &$+$       &$+$       &$+$    &$+$  &$+$  &$+$  &$+$  &$+$ &$+$ \\
\hline
\end{tabular}
\end{center}
\caption{$S_4\times Z^V_9\times Z^W_9\times Z^P_9 \times Z_{18}\times Z_2$ 
assignment of superfields
(Where the index  $i$ of the $S_4$ doublets runs $i=1,2$,
and the index $a$ of the $S_4$ triplets runs $a=1,2,3$.
The details of $S_4$ are given in Ref. \cite{s4}.)}
\end{table}

The superpotential defined in Eq.(5)-Eq.(9) has following problems. 
As the interaction $W_G$ induces too fast proton decay, they must be strongly suppressed. 
The mass parameter $M_\Phi$ in $W_\Phi$ must be forbidden in order to 
allow $U(1)_Z$ symmetry breaking.
In $W_0$, the contributions to flavor changing processes from the extra Higgs bosons must be
suppressed \cite{u1fv}. These problems should be solved by flavor symmetry.

If we introduce $S_4$ flavor symmetry and
assign $G,G^c$ to be triplets, then $W_G$ defined in Eq.(8) is forbidden. This is
because any products of doublets and singlets of $S_4$ do not contain triplets.
Note that we assume full $E_6$ symmetry does not realize at Planck scale,
therefore there is no need to assign all superfields to the same flavor representations.
In this model the generation number three is imprinted in $G, G^c$.
Therefore they may be called "G Higgs" (generation number imprinted colored Higgs).

As the existence of G Higgs which has life time longer than 0.1 second spoils the
success of Big Ban nucleosynthesis (BBN)\cite{reheating}, $S_4$ symmetry must be broken.
Therefore we assign $\Phi^c$ to be triplet and $\Phi$ to be doublet and singlet to forbid
$M_\Phi\Phi\Phi^c$.
With this assignment, $S_4$ symmetry is broken due to the VEV of $\Phi$ and the
effective trilinear terms are induced by non-renormalizable terms
\eqn{
W_{NRG}=\frac{1}{M^2_P}\Phi\Phi^c\L(GQQ+G^cU^cD^c+GU^cE^c+G^cQL+GD^cN^c\R).
}
The size of effective coupling constants of these terms should be
\eqn{
\frac{\L<\Phi\R>\L<\Phi^c\R>}{M^2_P}> 10^{-14},
}
to satisfy the BBN constraint \cite{f-extra-u1}.
The superpotential of gauge non-singlets $\Phi,\Phi^c$ is given by
\eqn{
W_\Phi&=&\frac{Y^\Phi_1}{M_P}\Phi^2_3[(\Phi^c_1)^2+(\Phi^c_2)^2+(\Phi^c_3)^2]
+\frac{\epsilon^2Y^\Phi_2}{M_P}[\Phi^2_1+\Phi^2_2]
[(\Phi^c_1)^2+(\Phi^c_2)^2+(\Phi^c_3)^2] \no \\
&+&\frac{\epsilon^2 Y^\Phi_3}{M_P}
[2\sqrt{3}\Phi_1\Phi_2((\Phi^c_2)^2-(\Phi^c_3)^2)
+(\Phi^2_1-\Phi^2_2)((\Phi^c_2)^2+(\Phi^c_3)^2-2(\Phi^c_1)^2)] \no \\
&+&\frac{\epsilon Y^\Phi_4}{M_P}\Phi_3[\sqrt{3}\Phi_1((\Phi^c_2)^2-(\Phi^c_3)^2)
+\Phi_2((\Phi^c_2)^2+(\Phi^c_3)^2-2(\Phi^c_1)^2)].
}
We assume  the global minimum of the potential $V(\Phi,\Phi^c)$ is at
$S_3$-symmetric vacuum as follow
\eqn{
\L<\Phi_1\R>=\L<\Phi_2\R>=0,\quad
\L<\Phi^c_1\R>=\L<\Phi^c_2\R>=\L<\Phi^c_3\R>=\frac{\L<\Phi_3\R>}{\sqrt{3}}
=\frac{V}{\sqrt{3}}.
}

The assignments of the other superfields are determined based on following criterion.
(1)The quark and lepton mass matrices reproduce observed mass hierarchies and
CKM and Maki-Nakagawa-Sakata 
(MNS) matrices. (2)The third generation Higgs $H^U_3,H^D_3$ are specified as MSSM Higgs
and the first and second generation Higgs superfields 
$H^{U,D}_{1,2}$ are almost inert Higgs (AIH). The representation of all superfields under the 
flavor symmetry is given in Table 2.

In order to realize Yukawa hierarchies, we introduce gauge singlet flavon superfields 
$V_i,X_W,W_i,P_W,X_P,P,X$
and fix the VEV of them by
\eqn{
&&\frac{\L<X\R>}{M_P}=\L(\frac{m_{SUSY}}{M_P}\R)^\frac{1}{16}=0.1=\epsilon , \quad
\frac{\L<X_W\R>}{M_P}=\frac{\L<X_P\R>}{M_P}
=\L(\frac{m_{SUSY}}{M_P}\R)^\frac{1}{7}=0.01=\epsilon^2 , \quad m_{SUSY}=O(\mbox{TeV}), \no \\
&&\frac{\L<W_1\R>}{M_P}=\epsilon^2\alpha c_W , \quad
\frac{\L<W_2\R>}{M_P}=\epsilon^2\beta s_W , \quad
\frac{\L<P\R>}{M_P}=\epsilon^2\gamma , \quad
\frac{\L<P_W\R>}{M_P}=\epsilon^2\gamma_W , \quad
\frac{\L<V_1\R>}{M_P}=\epsilon^2 c_V , \quad
\frac{\L<V_2\R>}{M_P}=\epsilon^2 s_V ,
}
where $c_V=\cos\theta_V, s_V=\sin\theta_V$ and Greek letters are complex
except for $\epsilon$.
In this paper, we assume original Lagrangian has CP symmetry and
all parameters in it are real. 
Therefore the complex VEVs given in Eq.(18) induce
spontaneous CP violation.
We illustrate how to get these VEVs as follows.
Without loss of generality, we can define $\L<X\R>,\L<X_W\R>,\L<X_P\R>,\L<V_1\R>$
to be real by the redefinition of superfields.
The superpotential of $V_i$ is given by
\eqn{
W=\frac{a}{M^6_P}(V^2_1+V^2_2)^3(3V^2_1V_2-V^3_2)
+\frac{b}{M^6_P}(3V^2_1V_2-V^3_2)^3.
}
This superpotential and soft SUSY breaking terms give polynomial potential as
\eqn{
V(s_3)=C_0+C_1s_3+ C_2s^2_3+C_4s^4_3, \quad s_3=\sin3\theta_V,
}
where all coefficients are real. The minimum condition of this potential with $\theta_V$,
\eqn{
c_3[C_1+2C_2s_3+4C_4s^3_3]=0,
}
has trivial solution $c_3=0$ and non-trivial solutions which satisfy the equation
\eqn{
C_1+2C_2s_3+4C_4s^3_3=0.
}
Which of these solutions is selected for the global minimum depends on
the parameters in the potential. 
Since the solution $c_3=0$ gives wrong prediction such as $V_{ud}=0$, 
we assume the solution of Eq.(22) corresponds to the global minimum. 
Whether the solution of Eq.(22) is real or complex also depends on
the parameters. As we choose real solution, $\L<V_2\R>$ has no relative phase.
On the other hand for $V(X_P,P)$ and $V(X_W,W_i,P_W)$, 
we assume $\L<W_1\R>,\L<W_2\R>,\L<P\R>,\L<P_W\R>$
have relative phases which break CP symmetry.
The scale of VEV is fixed as follows. If the superpotential of superfield $\Psi$
is given by
\eqn{
W=\frac{\Psi^n}{nM^{n-3}_P},
}
then the potential of $\Psi$ is given by
\eqn{
V(\Psi)=m^2_\Psi|\Psi|^2-\frac{A\Psi^n}{M^{n-3}_P}+\frac{|\Psi^{n-1}|^2}{M^{2n-6}_P},
}
where $m_\Psi\sim A \sim m_{SUSY}$ is assumed.
At the global minimum $\L<\Psi\R>\neq 0$, as the each terms in the potential should be balanced,
the scale of VEV is fixed by
\eqn{
\frac{\L<\Psi\R>}{M_P} \sim \L(\frac{m_{SUSY}}{M_P}\R)^{\frac{1}{n-2}}.
}

\subsection{Soft SUSY breaking terms}

We assume that the effect of SUSY breaking in hidden sector is
mediated by gravity and induces soft SUSY breaking terms in observable sector.
Since these terms are non-universal in general, large flavor-changing processes
are induced by the sfermion exchange. 
From the experimental constraints on them, the assignment of quarks and leptons
under the flavor symmetry are restrictive.
 
After the flavor violation, soft breaking scalar squared  mass matrices become
non-diagonal. For the Higgs scalars, this gives the mixing mass terms as follow
\eqn{
V=m^2\epsilon^4(H^U_3)^*H^U_i+
m^2\epsilon^5(H^D_3)^*H^D_i+m^2\epsilon^9S^*_3S_2,
}
which compel extra scalars to develop VEVs as
\eqn{
\L<H^U_i\R>=O(\epsilon^4)v_u ,\quad
\L<H^D_i\R>=O(\epsilon^5)v_d , \quad
\L<S_2\R>=O(\epsilon^9)v_s,
}
where we put
\eqn{
\L<H^U_3\R>=v_u=150.7\mbox{GeV} ,\quad
\L<H^D_3\R>=v_d=87.0\mbox{GeV} , \quad
\L<S_3\R>=v_s\gg \sqrt{v^2_u+v^2_d}=174.0\mbox{GeV} .
}
The VEVs of the extra scalar are very small, we call
$H^U_i$ and $H^D_i$ "almost inert-Higgs" (AIH) and $S_2$ as
"almost inert-singlet" (AIS). As $S_1$ does not develop VEV,
we call it "inert-singlet"(IS).
As the same effects affect the flavons, the VEV directions 
given in Eq.(17) and (18) are modified as follows,
\eqn{
\L<\Phi_1\R>\sim \L<\Phi_2\R>\sim O(\epsilon^4)V, \quad
\L<\Phi^c_a\R>=\frac{V}{\sqrt{3}}+O(\epsilon^4)V, \quad 
\frac{\L<V_1\R>}{M_P}=\epsilon^2c_V+O(\epsilon^6), \quad \cdots, 
}
and so on. Note that dominant parts of scalar squared mass matrices
of AIH and G Higgs are diagonal and degenerated.
Due to the smallness of VEVs of AIH and AIS, the superpartners
of AIH and G Higgs also have diagonal and degenerated mass matrices.
Therefore the trace of $S_4$ symmetry is imprinted in
their mass spectra which may be testable for LHC or future collider.

\subsection{The size of $V$}

The size of $V$ given in Eq.(17) should satisfy following condition.
The interactions of G Higgs are given by
\eqn{
W=\frac{1}{M^2_P}\Phi_3\Phi^c_aG_a(U^c_3E^c_3+Q_3Q_3)
=\frac{V^2}{\sqrt{3}M^2_P}(G_1+G_2+G_3)(U^c_3E^c_3+2U_3D_3),
}
from which each  G Higgs can decay to top and tau,
where we assume G Higgs is lighter than G higgsino and
G higgsino can decay to G Higgs.
The decay width of G Higgs is given by
\eqn{
\Gamma(G\to t+\tau)+\Gamma(G\to t+b)
\simeq \frac{m_G}{16\pi}\L(\frac{V^2}{\sqrt{3}M^2_P}\R)^2(5^2+8\times 13^2),
}
where $m_G$ is mass of G Higgs and the factor 5 and 13 are the approximate value of
renormalization factors \cite{p-s4u1}.
As the the life time of G Higgs
\eqn{
\tau(G)\simeq 7.2\times 10^{-29}\L(\frac{\mbox{TeV}}{m_G}\R)\L(\frac{M_P}{V}\R)^4\mbox{sec},
}
should be shorter than 0.1 second in order not to spoil the success of BBN, 
we have to require  
\eqn{
m_G>0.072 \L(\frac{10^{-6.5}M_P}{V}\R)^4\mbox{TeV}.
}
For $m_G=1$TeV, $V/M_P>0.52\times 10^{-6.5}$ should be satisfied.
Assuming the A-term dominant potential as
\eqn{
V(\Phi,\Phi^c)=\L\{-A\Phi^2_3[(\Phi^c_1)^2+(\Phi^c_2)^2+(\Phi^c_3)^2]+h.c.\R\}
+(D-\mbox{term})+(F-\mbox{term}),
}
the size of $V$ is given by
\eqn{
\frac{V}{M_P}\sim \frac{1}{Y^\Phi_1}\L(\frac{A}{M_P}\R)^\frac12
\sim (Y^\Phi_1)^{-1}\times 10^{-7.5},
}
which requires $Y^\Phi_1\sim 0.1$ to give $V/M_P=10^{-6.5}$.
Note that we assume $Y=O(1)$ means $0.1\leq Y \leq 1$ in this paper. 
Therefore the natural size of $V$ is given by $V/M_P\leq 10^{-6.5}$.
In order to satisfy Eq.(33) for TeV scale G Higgs,
the size of $V$ should be in the region as follow 
\eqn{
0.52\times 10^{-6.5} \leq \frac{V}{M_P}\leq 10^{-6.5}.
}


\section{Quark Sector}

The superpotential of quark sector is given by
\eqn{
W=H^U_3QY^UU^c+H^D_3QY^DD^c+H^U_iQY^U_{AIH}U^c+H^U_iQY^D_{AIH}D^c ,
}
where Yukawa matrices are calculated as follows,
\eqn{
Y^U&=&\Mat3{Y^U_1c_V\epsilon^6}{-Y^U_2s_V\epsilon^3}{Y^U_4c_V\epsilon^2}
{Y^U_1s_V\epsilon^6}{Y^U_2c_V\epsilon^3}{Y^U_4s_V\epsilon^2}
{Y^U_5\epsilon^4}{0}{Y^U_3}, \\
Y^D&=&\Mat3{-Y^D_1c_{2V}\epsilon^5}{Y^D_2s_{2V}\epsilon^4}
{Y^D_4(a_sc_V-a_p\gamma s_V)\epsilon^4}
{Y^D_1s_{2V}\epsilon^5}{Y^D_2c_{2V}\epsilon^4}
{Y^D_4(a_ss_V+a_p\gamma c_V)\epsilon^4}
{0}{0}{Y^D_3\epsilon^2},
}
where $c_{2V}\equiv \cos 2\theta_V$ and $s_{2V}\equiv \sin 2\theta_V$.
As the K\"ahler potential receives the effect of flavor violation,
superfields must be redefined as
\eqn{
&&U^c\to V_K(U)U^c , \quad
D^c\to V_K(D)D^c , \quad
Q\to V_K(Q)Q, \no \\
&&V_K(U)=\Mat3{1}{\alpha_1\epsilon^7}{u_2\epsilon^4}
{\alpha^*_1\epsilon^7}{1}{\alpha_3\epsilon^5}
{u_2\epsilon^4}{\alpha^*_3\epsilon^5}{1} , \quad
V_K(D)=\Mat3{1}{\beta_1\epsilon^5}{\beta_2\epsilon^9}
{\beta^*_1\epsilon^5}{1}{\beta_3\epsilon^8}
{\beta^*_2\epsilon^9}{\beta^*_3\epsilon^8}{1} , \no \\
&&V_K(Q)=\Mat3{1}{\gamma_1\epsilon^4}{q_1c_V\epsilon^2}
{\gamma^*_1\epsilon^4}{1}{q_1s_V\epsilon^2}
{q_1c_V\epsilon^2}{q_1s_V\epsilon^2}{1},
}
in order to get canonical kinetic terms \cite{kahler}.
As the result, quark mass matrices are given by
\eqn{
M'_U&=&V^T_K(Q)Y^Uv_uV_K(U)
=v_u\Mat3{Y^U_1c_V\epsilon^6}{-Y^U_2s_V\epsilon^3}{Y^U_4c_V\epsilon^2}
{Y^U_1s_V\epsilon^6}{Y^U_2c_V\epsilon^3}{Y^U_4s_V\epsilon^2}
{Y^U_5\epsilon^4}{\alpha^*_3Y^U_3\epsilon^5}{Y^U_3} , \\
M'_D&=&V^T_K(Q)Y^Uv_dV_K(D)
=v_d\Mat3{-Y^D_1c_{2V}\epsilon^5}{Y^D_2s_{2V}\epsilon^4}{Y^D_4\alpha_D\epsilon^4}
{Y^D_1s_{2V}\epsilon^5}{Y^D_2c_{2V}\epsilon^4}{Y^D_4\beta_D\epsilon^4}
{-q_1c_{3V}Y^D_1\epsilon^7}{q_1s_{3V}Y^D_2\epsilon^6}{Y^D_3\epsilon^2}, \\
&&\alpha_D=a_sc_V-a_p\gamma s_V ,\quad
\beta_D=a_ss_V+a_p\gamma c_V , \quad 
|\alpha_D|^2+|\beta_D|^2=1,
}
where some parameters are redefined for simplicity,
such as $Y^U_5+u_2Y^U_3\to Y^U_5$.
Note that the each elements in these matrices include only
leading terms and the contributions to mass matrices from $Y_{AIH}$ are negligible
with this approximation.
These matrices are diagonalized by the superfield redefinitions
\eqn{
&&U \to L_UU , \quad D\to L_DD , \quad U^c \to R_UU^c , \quad D^c \to R_DD^c , \\
L^T_U&=&\Mat3{c_V}{s_V}{-(Y^U_4/Y^U_3)\epsilon^2}
{-s_V}{c_V}{0}
{(Y^U_4/Y^U_3)c_V\epsilon^2}{(Y^U_4/Y^U_3)s_V\epsilon^2}{1}, \\
L^T_D&=&\Mat3{c_{2V}}{-s_{2V}}{-(Y^D_4/Y^D_3)(\alpha_D c_{2V}-\beta_D s_{2V})\epsilon^2}
{s_{2V}}{c_{2V}}{-(Y^D_4/Y^D_3)(\alpha_D s_{2V}+\beta_D c_{2V})\epsilon^2}
{(Y^D_4/Y^D_3)\alpha^*_D\epsilon^2}{(Y^D_4/Y^D_3)\beta^*_D\epsilon^2}{1}, \\
R_U&=&\Mat3{1}{N_U\alpha^*_3\epsilon^7}{(Y^U_5/Y^U_3)\epsilon^4}
{-N_U\alpha_3\epsilon^7}{1}{\alpha_3\epsilon^5}
{-(Y^U_5/Y^U_3)\epsilon^4}{-\alpha^*_3\epsilon^5}{1}, \\
R_D&=&\Mat3{1}{\xi_3\epsilon^5}
{(\xi^*_1/Y^D_3)\epsilon^5}
{-\xi^*_3\epsilon^5}{1}{(\xi^*_2/Y^D_3)\epsilon^4}
{-(\xi_1/Y^D_3)\epsilon^5}
{-(\xi_2/Y^D_3)\epsilon^4}{1},
}
from which we get
\eqn{
L^T_UM'_UR_U&=&\mbox{diag}(m_u,m_c,m_t)
=\mbox{diag}\L((Y^U_1-Y^U_4Y^U_5/Y^U_3)\epsilon^6v_u,Y^U_2\epsilon^3v_u,Y^U_3v_u\R), \\
L^T_DM'_DR_D&=&\mbox{diag}(m_d,m_s,m_b)
=\mbox{diag}\L(-Y^D_1\epsilon^5v_d , Y^D_2\epsilon^4v_d ,Y^D_3\epsilon^2v_d \R), \\
V_{CKM}&=&L^\dagger_UL_D
=\Mat3{c_{3V}}{s_{3V}}{V_{ub}}
{-s_{3V}}{c_{3V}}{V_{cb}}
{V_{td}}{V_{ts}}{1} , \\
&&V_{ub}=(r_D(\alpha^*_Dc_V+\beta^*_Ds_V)-r_U)\epsilon^2
=(a_sr_D-r_U)\epsilon^2, \no \\
&&V_{cb}=r_D(\beta^*_Dc_V-\alpha^*_Ds_V)\epsilon^2
=a_pr_D\gamma^*\epsilon^2 , \no \\
&&V_{td}=(r_Uc_{3V}-r_D(\alpha_D c_{2V}-\beta_D s_{2V}))\epsilon^2
=[c_{3V}(r_U-a_sr_D)+a_pr_Ds_{3V}\gamma]\epsilon^2 , \no \\
&&V_{ts}=(r_Us_{3V}-r_D(\alpha_D s_{2V}+\beta_D c_{2V}))\epsilon^2
=[s_{3V}(r_U-a_sr_D)-a_pr_Dc_{3V}\gamma]\epsilon^2 , \no \\
&&r_U=Y^U_4/Y^U_3 , \quad r_D=Y^D_4/Y^D_3.
}
The experimental values of CKM matrix and quark running masses at 1TeV
\eqn{
&&\Mat3{|V_{ud}|}{|V_{us}|}{|V_{ub}|}
{|V_{cd}|}{|V_{cs}|}{|V_{cv}|}
{|V_{td}|}{|V_{ts}|}{|V_{tb}|}
=\Mat3{0.974}{0.225}{0.355\times 10^{-2}}
{0.225}{0.973}{4.14\times 10^{-2}}
{0.886\times 10^{-2}}{4.05\times 10^{-2}}{1} , \no \\
&&J=\mbox{Im}(V_{us}V_{cb}V^*_{ub}V^*_{cs})=3.06\times 10^{-5} \quad \cite{PDG2012} , \\
&&m_u=(1.17 \pm 0.35)\times 10^{-3} ,\quad 
m_c=0.543^{+0.037}_{-0.072} ,\quad m_t=148.1\pm 1.3 , \no \\
&&m_d=(2.40^{+0.42}_{-0.41})\times 10^{-3} ,\quad 
m_s=(4.9^{+1.5}_{-1.0})\times 10^{-2} ,\quad m_b=2.41^{+0.14}_{-0.05}  \quad (\mbox{GeV}) 
\quad \cite{mass},  
}
are realized by putting the parameters at $\mu=M_P$ as follows
\eqn{
&&r_U-a_sr_D=0.355 , \quad a_pr_D=4.14 , \quad \mbox{arg}(\gamma)=108.379^\circ , \quad
3\theta_V=13.003^\circ , \no \\
&&\L|Y^U_1-Y^U_4Y^U_5/Y^U_3\R|=1.5, \quad 
\L|Y^U_2\R|=0.71 , \quad \L|Y^U_3\R|=0.28 , \no \\
&&\L|Y^D_1\R|=0.38 , \quad \L|Y^D_2\R|=0.78 , \quad 
\L|Y^D_3\R|=0.38 ,
}
where we use the renormalization factors given in \cite{p-s4u1}.
As these parameters are consistent with the assumption that all factors $Y$ are $O(1)$,
quark mass hierarchy is realized without fine tuning.    
The soft SUSY breaking squared mass matrices of squarks are given by
\eqn{
\frac{m^2_U}{m^2}=\Mat3{O(1)}{\alpha_1\epsilon^7}{\epsilon^4}
{\alpha^*_1\epsilon^7}{O(1)}{\alpha_2\epsilon^5}
{\epsilon^4}{\alpha^*_2\epsilon^5}{O(1)} , 
\frac{m^2_D}{m^2}=\Mat3{O(1)}{\beta_1\epsilon^5}{\beta_2\epsilon^9}
{\beta^*_1\epsilon^5}{O(1)}{\beta_3\epsilon^8}
{\beta^*_2\epsilon^9}{\beta^*_3\epsilon^8}{O(1)} , 
\frac{m^2_Q}{m^2}=\Mat3{1}{\gamma_1\epsilon^4}{c_V\epsilon^2}
{\gamma^*_1\epsilon^4}{1}{s_V\epsilon^2}
{c_V\epsilon^2}{s_V\epsilon^2}{O(1)}, 
}
and the squark A-term matrices are given by
\eqn{
V&\supset& -v_uUA_UU^c-v_dDA_DD^c+h.c. , \\
A_U&=&\Mat3{A^U_1c_V\epsilon^6}{-A^U_2s_V\epsilon^3}{A^U_4c_V\epsilon^2}
{A^U_1s_V\epsilon^6}{A^U_2c_V\epsilon^3}{A^U_4s_V\epsilon^2}
{A^U_5\epsilon^4}{\alpha^*_3A^U_3\epsilon^5}{A^U_3} , \\
A_D&=&
\Mat3{-A^D_1c_{2V}\epsilon^5}{A^D_2s_{2V}\epsilon^4}{A^D_4\alpha'\epsilon^4}
{A^D_1s_{2V}\epsilon^5}{A^D_2c_{2V}\epsilon^4}{A^D_4\beta'\epsilon^4}
{-q_1c_{3V}A^D_1\epsilon^7}{q_1s_{3V}A^D_2\epsilon^6}{A^D_3\epsilon^2},
}
where these matrices are defined for canonically normalized superfields.
After the diagonalization of Yukawa matrices, the squared mass matrices are given by
\eqn{
(m^2_U)_{SCKM}&=&R^\dagger_Um^2_UR_U
=m^2\Mat3{O(1)}{\epsilon^7}{\epsilon^4}
{\epsilon^7}{O(1)}{\epsilon^5}
{\epsilon^4}{\epsilon^5}{O(1)} ,\\
(m^2_D)_{SCKM}&=&R^\dagger_Dm^2_UR_D
=m^2\Mat3{O(1)}{\epsilon^5}{\epsilon^5}
{\epsilon^5}{O(1)}{\epsilon^4}
{\epsilon^5}{\epsilon^4}{O(1)} ,\\
(m^2_Q)_{SCKM}&=&L^\dagger_{U,D}m^2_QL_{U,D}
=m^2\Mat3{1}{\epsilon^4}{\epsilon^2}
{\epsilon^4}{1}{\epsilon^2}
{\epsilon^2}{\epsilon^2}{O(1)},
}
and the (1,1) elements of $A_U$ and $A_D$ remain real except for
the next leading terms which are at most $O(\epsilon^{10})$ and $O(\epsilon^{9})$,
therefore the SUSY contributions to electric dipole moment of the neutron are negligible.
Note that the (1,2) and (2,1) elements of $(m^2_Q)_{SCKM}$ are complex.
The off-diagonal elements of squark mass matrices contribute to
flavor and CP violation through the squark exchange, on which are imposed
severe constraints.
With the mass insertion approximation, the most stringent bound for the squark mass $M_Q$
is given by $\epsilon_K$ as
\eqn{
\sqrt{\frac{\mbox{Im}[(m^2_Q)_{12}(m^2_D)_{12}]}{M^4_Q}}=
\epsilon^{4.5}&<&4.4\times 10^{-4}\L(\frac{M_Q}{\mbox{TeV}}\R) 
\quad\to\quad M_Q> 72\mbox{GeV},
}
where $M_Q=M(\mbox{gluino})=M(\mbox{squark})$ is assumed \cite{SUSYFCNC}.
This bound is very weak and the SUSY FCNC problem is solved.
Note that if CP symmetry was not imposed, $A^D_1$ would be complex and
the constraint on neutron EDM would gave stronger bound, $M_Q>625$GeV.

The contribution to FCNC from AHI exchange is also suppressed due to the
hierarchical structure of $Y^{U,D}_{AIH}$.
Assuming the mass degeneracy of CP-even AIH and CP-odd AIH,
the strongest mass bound for AIH is given by $D^0-\bar{D}^0$ as
$m_{AIH}>79$GeV \cite{u1fv}.


\section{Lepton Sector}

The superpotential of lepton sector is given by
\eqn{
W=H^D_3LY^EE^c+H^U_3LY^NN^c+\frac12\Phi_3N^cY^M N^c
+H^D_iLY^E_{AIH}E^c+H^U_iLY^N_{AIH}N^c+\frac12\Phi_iN^cY^M_i N^c ,
}
where
\eqn{
Y^E&=&\Mat3{Y^E_1\epsilon^5}{0}{0}
{Y^E_4\alpha_E\epsilon^5}
{-Y^E_2\beta s_W\epsilon^3}{Y^E_3\alpha c_W\epsilon^2}
{Y^E_4\beta_E\epsilon^5}
{Y^E_2\alpha c_W\epsilon^3}{Y^E_3\beta s_W\epsilon^2}, \\
Y^N&=&\epsilon^5\Mat3{Y^N_1c_V}{Y^N_1s_V}{0}
{Y^N_2+Y^N_5\beta s_W}{Y^N_5\alpha c_W-Y^N_4\gamma_W}{0}
{Y^N_5\alpha c_W+Y^N_4\gamma_W}{Y^N_2-Y^N_5\beta s_W}{0} ,\\
Y^M&=&\Mat3{Y^M_1\epsilon^6}{0}{0}
{0}{Y^M_1\epsilon^6}{0}
{0}{0}{Y^M_3\epsilon^4}, \\
&&\alpha_E=b_sc_V+b_w(\beta c_Vs_W+\alpha s_Vc_W)-b_p\gamma_W s_V , \no \\
&&\beta_E=b_ss_V+b_w(\alpha c_Vc_W-\beta s_Vs_W)+b_p\gamma_W c_V , \quad
|\alpha_E|^2+|\beta_E|^2=1 .
}
Because the all elements of MNS matrix are $O(1)$,
the contributions of flavor violation in K\"ahler potential and $Y_{AIH}$ and 
$Y^M_i$ are negligible. Therefore the lepton mass matrices are given by
\eqn{
M_E&=&Y^Ev_d=v_d\Mat3{Y^E_1\epsilon^5}{0}{0}
{Y^E_4\alpha_E\epsilon^5}{-Y^E_2\beta s_W\epsilon^3}{Y^E_3\alpha c_W\epsilon^2}
{Y^E_4\beta_E\epsilon^5}{Y^E_2\alpha c_W\epsilon^3}{Y^E_3\beta s_W\epsilon^2} , \\
\sqrt{M_\nu}&=&\frac{Y^Nv_u}{\sqrt{(Y^M)_{11}V}}=\Mat3{x_1c_V}{x_1s_V}{0}
{x_2+x_5\beta s_W}{x_5\alpha c_W-x_4\gamma_W}{0}
{x_5\alpha c_W+x_4\gamma_W}{x_2-x_5\beta s_W}{0} , \\
&&x_i=\frac{Y^N_i\epsilon^2v_u}{\sqrt{Y^M_1V}}  \quad (i=1,2,4,5).
}
The charged lepton mass matrix is diagonalized by superfield redefinitions
\eqn{
&&E\to L_EE , \quad E^c\to R_EE^c  ,\quad N\to L_EN \\
L^T_E&=&\Mat3{1}{\epsilon^4}{\epsilon^6}
{\epsilon^4}{1}{\epsilon^2}
{\epsilon^6}{\epsilon^2}{1}
\Mat3{1}{0}{0}
{0}{\beta s_W}{-\alpha c_W}
{0}{\alpha^*c_W}{\beta^*s_W} , \\
R_E&=&\Mat3{1}{\epsilon^2}{\epsilon^3}
{\epsilon^2}{1}{\epsilon}
{\epsilon^3}{\epsilon}{1},
}
from which we get
\eqn{
L^T_EM_ER_E&=&\mbox{diag}(m_e,m_\mu,m_\tau)
=\mbox{diag}\L(Y^E_1\epsilon^5v_d , 
-Y^E_2(\beta^2s^2_W+\alpha^2c^2_W)\epsilon^3v_d , Y^E_3\epsilon^2v_d\R) , \\
M'_\nu&=&L^T_E\sqrt{M_\nu}\sqrt{M_\nu}^TL_E \no \\
&=&\Mat3{x^2_1}{x_1(M_{21}c_V+M_{22}s_V)}{x_1(M_{31}c_V+M_{32}s_V)}
{x_1(M_{21}c_V+M_{22}s_V)}{M^2_{21}+M^2_{22}}{M_{21}M_{31}+M_{22}M_{32}}
{x_1(M_{31}c_V+M_{32}s_V)}{M_{21}M_{31}+M_{22}M_{32}}{M^2_{31}+M^2_{32}}, \\
&&M_{21}=\beta s_Wx_2
+(\beta^2s^2_W-\alpha^2c^2_W)x_5-\alpha\gamma_W c_Wx_4 , \\
&&M_{31}=\alpha^*c_Wx_2 ,
+(\alpha^*\beta+\alpha\beta^*)c_Ws_Wx_5+\beta^*\gamma_W s_Wx_4 , \\
&&M_{22}=-\alpha c_W x_2+2\alpha\beta c_Ws_Wx_5-\beta\gamma_W s_Wx_4 , \\
&&M_{32}=\beta^* s_Wx_2
+(c^2_W-s^2_W)x_5-\alpha^*\gamma_W c_Wx_4.
}
The experimental values of charged lepton running masses at 1TeV
\eqn{
m_e=4.895\times 10^{-4}, \quad m_\mu=0.1033, \quad m_\tau=1.757, \quad
(\mbox{GeV}) \quad \cite{mass},
}
are realized by putting the parameters at $\mu=M_P$ as follows
\eqn{
\L|Y^E_1\R|=0.30 , \quad \L|Y^E_2(\alpha^2c^2_W+\beta^2s^2_W)\R|=0.62 , \quad
\L|Y^E_3\R|=1.1 , 
}
where we use the renormalization factors given in \cite{p-s4u1}.
Charged lepton mass hierarchy is realized without fine tuning which
is the same as for quark sector.

In order to realize neutrino mass scale $m_\nu\sim O(0.01)\mbox{eV}$, the relations
\eqn{
x^2_i = \frac{(Y^N_i\epsilon^2 v_u)^2}{Y^M_1V}\sim O(0.01)\mbox{eV}
\quad\to\quad Y^M_1V \sim 10^{11} \mbox{GeV} , \quad
M_1=\epsilon^6Y^M_1V \sim 10^{5} \mbox{GeV}
}
are required.
After the diagonalization of charged lepton Yukawa matrix,
the squared mass matrix and A-term matrix are given by
\eqn{
(m^2_L)_{SMNS}&=&(L^T_Em^2_LL^*_E)=m^2\Mat3{O(1)}{\epsilon^4}{\epsilon^4}
{\epsilon^4}{1}{\epsilon^4}
{\epsilon^4}{\epsilon^4}{1}, \\
(A'_E)_{SMNS}&=&L^T_EA_ER_E
=v_d\Mat3{A_1\epsilon^5}{O(\epsilon^{7})}{O(\epsilon^{6})}
{O(\epsilon^5)}{O(\epsilon^{3})}{O(\epsilon^{4})}
{O(\epsilon^5)}{O(\epsilon^{3})}{O(\epsilon^2)},
}
where the (1,1) element of $A'_E$ is real at leading order 
and the SUSY contribution to
electric dipole moment of the electron is negligible. 
Based on consideration of the lepton flavor violations,  
the most stringent bound for slepton mass $M_L$ is given by
$\mu\to e+\gamma$ as
\eqn{
\frac{v_d}{M_L}\epsilon^5<3.4\times 10^{-6}
\L(\frac{M_L}{300\mbox{GeV}}\R) \quad\to\quad
M_L> 300\mbox{GeV},
}
where $M_L=M(\mbox{slepton})=M(\mbox{photino})$ is assumed \cite{SUSYFCNC}.
Without CP symmetry, the constraint on electron EDM would give stronger bound,
$M_L>1765$GeV.

The contribution to lepton flavor violation from AHI exchange is also suppressed due to the
hierarchical structure of $Y^E_{AIH}$ as same as quark sector.
Assuming the mass degeneracy of CP-even AIH and CP-odd AIH,
the strongest mass bound for AIH is given by $\mu\to e+\gamma$ as
$m_{AIH}>38$GeV \cite{u1fv}.

For the canonically normalized superfields, RHN mass matrix is given by
\eqn{
M_N&=&V^T_K(N)Y^MVV_K(N)=\Mat3{M_1}{M\epsilon^4}{0}
{M\epsilon^4}{M_1(1+\epsilon^4)}{0}
{0}{0}{M_3} , 
\quad M\sim M_1 , \quad M_3=Y^N_3V\epsilon^4, \\
&&V_K(N)=\Mat3{1}{\epsilon^4}{0}
{\epsilon^4}{1}{0}
{0}{0}{1},
}
which gives degenerated mass spectrum of RHNs as follow
\eqn{
M_1\simeq M_2=M_1(1+\epsilon^4) \ll  M_3 \quad \to \quad
\delta_N=\frac{M_2-M_1}{M_1}\sim \epsilon^4 .
}
The right handed sneutrinos have same spectrum.
In the early universe, the out-of-equilibrium decay of $n^c_1$ and $N^c_1$
generates B-L asymmetry which is transferred to a baryon asymmetry
by EW sphaleron processes. Following Ref. \cite{leptogenesis},
the baryon asymmetry is given by
\eqn{
B_f&\sim& -\frac{\kappa\epsilon_{CP}}{3g_*},
}
where $g_*=340$ is the degree of freedom of radiation,
$\kappa$ is dilution factor which is given by
\eqn{
\kappa&\sim&\frac{1}{K\ln K} , \\
&&K=\frac{\Gamma(M_1)}{2H(M_1)}, \quad
\Gamma(M_1)=\frac{K_{11}M_1}{8\pi}, \quad
H(M_1)=\sqrt{\frac{\pi^2g_*M^4_1}{90M^2_P}}, \quad
K_{ij}=\sum^3_{l=1}(Y^N_{li})^*(Y^N_{lj}),
}
and $\epsilon_{CP}$ is given by
\eqn{
\epsilon_{CP}=-\frac{1}{2\pi}\frac{\mbox{Im}(K^2_{12})}{K_{11}}
\L(\frac{2\sqrt{x}}{x-1}+\sqrt{x}\ln\frac{1+x}{x}\R)
\simeq -\frac{\mbox{Im}[K^2_{12}]}{2\pi K_{11}\delta_N}, \quad
x=\frac{M^2_2}{M^2_1}\simeq 1+2\delta_N.
}
From the order estimations as follows
\eqn{
K_{12}\sim K_{11} \sim \epsilon^{10}, \quad 
K\sim \L(\frac{0.7\mbox{PeV}}{M_1}\R), \quad
\epsilon_{CP}\sim 10^{-6} , \quad
M_1\sim 10^5\mbox{GeV},
}
we get the correct amount of baryon asymmetry $B_f\sim 10^{-10}$.
Right sign of baryon number corresponds to
\eqn{
0^\circ <2\theta_{CP}<180^\circ, \quad \theta_{CP}=\mbox{arg}(K_{12}) .
}


\section{Dark Matter}

The lightest SUSY particles of this model is singlino
$s_2$ which is the superpartner of AIS $S_2$.
The superpotential of Higgs sector is given by
\eqn{
W&=&\lambda_1\epsilon S_3(H^U_1H^D_1+H^U_2H^D_2)+\lambda_3S_3H^U_3H^D_3 \no \\
&+&\lambda_2\epsilon^6S_2[c_V(H^U_1H^D_2+H^U_2H^D_1)+s_V(H^U_1H^D_1-H^U_2H^D_2)] \no \\
&+&\lambda_4\epsilon^5(c_VH^U_1+s_VH^U_2)S_2H^D_3 
+\lambda_5\epsilon^6H^U_3S_2(c_VH^D_1+s_VH^D_2),
}
which gives 
\eqn{
M(s_2)\sim \frac{(\epsilon^5\lambda_4v_d)(\epsilon^6\lambda_5)v_u}{\epsilon\lambda_1v_s}
\sim 1\mbox{eV}.
} 
Although $s_2$ is not the dominant component of dark matter, it may
help to explain the delay of structure formation \cite{cluster}.
The massless singlino $s_1$ and LSP $s_2$ behave as extra neutrinos
and change the effective neutrino generation number to
\eqn{
N_{\mbox{eff}}=3.194 \quad \cite{massless-singlino},
}
where $m_{Z'}<4700$GeV is assumed. This extra contributions soften
the discrepancy of the expansion rate $H_0$ between
the measurements of type Ia supernovas and Cepheid variable and
CMB data \cite{cluster}.

The interaction of bino is given by
\eqn{
{\cal L}=-i\frac{g_Y}{\sqrt{2}}(H^D_3)^*\lambda_Yh^D_3,
}
where higgsino $h^D_3$ has mixing mass term with $s_2$ as follow
\eqn{
{\cal L}=\lambda_4\epsilon^5\L<H^U_i\R>s_2h^D_3\sim \epsilon^9v_us_2h^D_3.
}
Therefore bino life time is calculated as follow
\eqn{
\Gamma(\lambda_Y\to H+s_2)
\sim\frac{g^2_Ym_{SUSY}}{4\pi}\L(\frac{\epsilon^9v_u}{m_{SUSY}}\R)^2
\sim 10^{-10}\mbox{eV} \quad\to\quad
\tau\sim 10^{-5}\mbox{sec},
}
which is consistent with standard cosmology.

Five of the six flavon multiplets $\Phi,\Phi^c$ have lighter masses than 100TeV
and are produced non-thermally  through the $U(1)_Z$ gauge interaction.
As the lightest flavon (LF) is quasistable, therefore it is the candidate of DM.
Solving the Boltzmann equation with the boundary condition $n_{LF}(T_{RH})=0$,
we get a relic abundance of the LF as \cite{u1pamela}
\eqn{
\Omega_{LF}h^2=5.0\times 10^{-3}\L(\frac{T_{RH}}{10^7\mbox{GeV}}\R)^3
\L(\frac{10^{12}\mbox{GeV}}{V}\R)^4\L(\frac{m_{LF}}{4\mbox{TeV}}\R).
}
The heaviest RHN $n^c_3$ behaves like LF and  has a relic abundance given by
\eqn{
\Omega_Nh^2=0.6\L(\frac{T_{RH}}{10^7\mbox{GeV}}\R)^3
\L(\frac{10^{12}\mbox{GeV}}{V}\R)^4\L(\frac{M_3}{10^7\mbox{GeV}}\R),
}
where $M_3$ is defined by Eq.(87) and we put $M_3\sim 10^7$GeV and $Y^M_3\sim 0.1$
based on following reason.
As $M_3\gg m_{LF}$ is satisfied, the contribution of LF to $\Omega_{DM}$ 
is negligible.
$n^c_3$ can decay through the interaction in K\"ahler potential as
\eqn{
K=\frac{1}{M^5_P}\L[X^*N^c_3(N^c_i)^*X_WW_iS_1(S_3)^*+(P_W\mbox{-contribution})\R]_{\theta^4}
=\frac{\epsilon^5}{M^2_P}n^c_3(\xi_1\bar{n}^c_1
+\xi_2\bar{n}^c_2)s_1\bar{s}_3,
}
and has a life time given by
\eqn{
\Gamma(n^c_3\to n^c_i+s_1+s_3)\sim\frac{\epsilon^{10}M^5_3}{O(100)\pi^3M^4_P} 
\sim 10^{-43}\mbox{eV} \quad\to\quad
\tau \sim 10^{28}\mbox{sec}.
}
The daughter particles $n^c_{1,2}$ decay to $\nu+H$ and
give neutrino flux at $E_\nu\sim M_3/6\sim \mbox{PeV}$.
From the latest results of IceCube \cite{icecube-exp}, the mass and life time of DM
are around PeV and $10^{28}$sec \cite{icecube-DM}.
If we identify $n^c_3$ as DM, $M_3$ is fixed at $10^7$GeV from the IceCube results.
Note that reheating temperature should be 
$T_{RH} < 10^7$GeV  to avoid gravitino over production.
After the three parameters in Eq.(103) are fixed as follows
\eqn{
0.52\times 10^{-6.5}<\frac{V}{M_P}<10^{-6.5} , \quad M_3\sim 10^7\mbox{GeV}, \quad
10^7\mbox{GeV}>T_{RH}>M_3,
}
the right amount of DM, $\Omega_Nh^2\sim O(1)-O(10)$ is realized.
This is an interesting prediction of our model.
For the reheating temperature $T_{RH}\sim 10^7$GeV, the thermal mass of flavon $\Phi_3$
is estimated as follow
\eqn{
m_T(\Phi_3)\sim \epsilon^6T_{RH}\sim 10 \mbox{GeV},
}
which is small enough to avoid symmetry restoration.
If inflaton $\Psi$ decays through Planck suppressed interaction, for example, such as
\eqn{
W=\frac{1}{M_P}\Psi H^U_3Q_3U^c_3,
}
then reheating temperature is given by
\eqn{
T_{RH}\sim \sqrt{M_P\Gamma(\Psi)}\sim \L(M_P\frac{m^3(\Psi)}{M^2_P}\R)^\frac12,
}
which requires  inflaton mass $m(\Psi)\sim 10^{11}$GeV for our model.

\subsection{Numerical analysis}

Finally we calculate detection probabilities of neutrino flavors emitted through
dark matter decay and argument  of $K_{12}$. 
The neutrino mass matrix is given by
\eqn{
(M'_\nu)_{exp}&=&U^*_{MNS}\mbox{diag}(0,m_2,m_3)U^\dagger_{MNS}, \\
U_{MNS}&=&\Mat3{c_{12}c_{13}}{s_{12}c_{13}}{s_{13}e^{-i\delta}}
{-s_{12}c_{23}-c_{12}s_{23}s_{13}e^{i\delta}}
{c_{12}c_{23}-s_{12}s_{23}s_{13}e^{i\delta}}{s_{23}c_{13}}
{s_{12}s_{23}-c_{12}c_{23}s_{13}e^{i\delta}}
{-c_{12}s_{23}-s_{12}c_{23}s_{13}e^{i\delta}}
{c_{23}c_{13}}
\mbox{diag}(1,e^{i\phi},1),
}
on the basis that charged lepton mass matrix is diagonalized.
From the experimental values as follows \cite{PDG2012},
\eqn{
\sin^2\theta_{12}&=&0.308 \pm 0.017 ,\\
\sin^2\theta_{23}&=&0.437^{+0.033}_{-0.023} ,\\
\sin^2\theta_{13}&=&0.0234^{+0.0020}_{-0.0019} ,\\
\delta/\pi&=&1.39^{+0.38}_{-0.27}, \\
\Delta m^2_{21}&=&(0.753\pm 0.018)\times 10^{-4}\mbox{eV}^2 ,\\
\Delta m^2_{32}&=&(24.4\pm 0.6) \times 10^{-4}\mbox{eV}^2 ,
}
we fix the parameters as follows
\eqn{
\theta_{12}&=&33.709^\circ, \\
\theta_{23}&=&41.381^\circ, \\
\theta_{13}&=& 8.799^\circ, \\
m_1&=&0 \mbox{eV} , \\
m_2&=&0.867756 \times 10^{-2}\mbox{eV} ,\\
m_3&=&5.015277 \times 10^{-2}\mbox{eV} ,\\
\delta&=&250.0^\circ.
}
By the phase rotations of lepton doublets $L_a$, we define
the diagonal elements of $(M'_\nu)_{exp}$ to be real and non-negative
and the real parts of (1,2) and (1,3) elements of $(M'_\nu)_{exp}$ to be non-negative.
The same procedure is performed for $M'_\nu$ which is given in Eq.(76).
The matching condition $M'_\nu=(M'_\nu)_{exp}$ gives seven equations
with nine free parameters as follows
\eqn{
\phi , \quad \theta_a=\mbox{arg}(\alpha) , \quad
\theta_b=\mbox{arg}(\beta) , \quad \theta_c=\mbox{arg}(\gamma_W) , \quad
\theta_W, \quad x_1 , \quad x_2 , \quad x_4 , \quad x_5.
}
Note that two of nine equations in matching condition
are automatically solved
due to the condition $\mbox{det}M'_\nu=\mbox{det}(M'_\nu)_{exp}=0$.
In order to solve the equations, two constraints should be added by hand.
We fix $(\phi,x_4)$ to solve the equations numerically 
and calculate branching ratios and argument of $K_{12}$.
The former is given by
\eqn{
B_e&=&\frac{\Gamma(n_1\to\nu_eH)+\Gamma(n_2\to\nu_eH)}{\Gamma(\nu H)}
=\frac{x^2_1}{x^2_1+2(x^2_2+x^2_5+x^2_4)} ,\\
B_\mu&=&\frac{\Gamma(n_1\to\nu_\mu H)+\Gamma(n_2\to\nu_\mu H)}{\Gamma(\nu H)}
=\frac{x^2_2+x^2_5+x^2_4+2x_2x_5s_W\cos\theta_b-2x_4x_5c_W\cos(\theta_a-\theta_c)}
{x^2_1+2(x^2_2+x^2_5+x^2_4)} ,\\
B_\tau&=&\frac{\Gamma(n_1\to\nu_\tau H)+\Gamma(n_2\to\nu_\tau H)}{\Gamma(\nu H)}
=\frac{x^2_2+x^2_5+x^2_4-2x_2x_5s_W\cos\theta_b+2x_4x_5c_W\cos(\theta_a-\theta_c)}
{x^2_1+2(x^2_2+x^2_5+x^2_4)} , \\
\Gamma(\nu H)&=&\Gamma(n_1\to\nu_eH)+\Gamma(n_2\to\nu_eH)
+\Gamma(n_1\to\nu_\mu H)+\Gamma(n_2\to\nu_\mu H) \no \\
&+&\Gamma(n_1\to\nu_\tau H)+\Gamma(n_2\to\nu_\tau H),
}
where we put $\Gamma(n_3\to n_1+s_1+s_3)=\Gamma(n_3\to n_2+s_1+s_3)$ by hand.
The latter is given by
\eqn{
\theta_{CP}=\mbox{arg}(K_{12})
&=&\mbox{arg}\L[
(x^2_1c_Vs_V+2x_2x_5c_W\cos\theta_a-2x_4x_5s_W\cos(\theta_b-\theta_c) )\R. \no \\
&+&\L. i(2x^2_5c_Ws_W\sin(\theta_a-\theta_b)-2x_2x_4\sin\theta_c)\R] ,
}
where we assume the contributions from AIH are negligible for simplicity.
Due to the neutrino oscillation, the neutrinos emitted by RHN-decay 
change the flavor, therefore the detection probabilities of neutrino at detector
are given by
\eqn{
P(\nu_e)&=&B_e(|U_{e1}|^4+|U_{e2}|^4+|U_{e3}|^4) \no \\
&+&B_\mu(|U_{e1}|^2|U_{\mu 1}|^2+|U_{e2}|^2|U_{\mu 2}|^2+|U_{e3}|^2|U_{\mu 3}|^2) \no \\
&+&B_\tau(|U_{e1}|^2|U_{\tau 1}|^2+|U_{e2}|^2|U_{\tau 2}|^2+|U_{e3}|^2|U_{\tau 3}|^2) ,\\
P(\nu_\mu)&=&B_e(|U_{e1}|^2|U_{\mu 1}|^2+|U_{e2}|^2|U_{\mu 2}|^2
+|U_{e3}|^2|U_{\mu 3}|^2) \no \\
&+&B_\mu(|U_{\mu 1}|^4+|U_{\mu 2}|^4+|U_{\mu 3}|^4) \no \\
&+&B_\tau(|U_{\mu 1}|^2|U_{\tau 1}|^2+|U_{\mu 2}|^2|U_{\tau 2}|^2
+|U_{\mu 3}|^2|U_{\tau 3}|^2) ,\\
P(\nu_\tau)&=&B_e(|U_{e1}|^2|U_{\tau 1}|^2+|U_{e2}|^2|U_{\tau 2}|^2
+|U_{e3}|^2|U_{\tau 3}|^2) \no \\
&+&B_\mu(|U_{\mu 1}|^2|U_{\tau 1}|^2+|U_{\mu 2}|^2|U_{\tau 2}|^2
+|U_{\mu 3}|^2|U_{\tau 3}|^2) \no \\
&+&B_\tau(|U_{\tau 1}|^4+|U_{\tau 2}|^4+|U_{\tau 3}|^4),
}
where we assume the baseline is longer than $\mbox{PeV}/m^2_\nu \sim 10^{-4}$pc.
As the equation $3\theta_V=13.003^\circ$ has
three solutions,
\eqn{
\theta_V=4.334^\circ, \quad 124.334^\circ, \quad 
244.334^\circ, 
}
we calculate $\theta_{CP}, P(\nu_{e,\mu,\tau})$ for each cases.
The results are given in Table 3.
The dependences of detection probability $P(\nu_{e,\mu,\tau})$ on parameters are weak.
Depending on $\phi$, both sign of $\sin(2\theta_{CP})$ are possible.

\begin{table}[htbp]
\begin{center}
$\theta_V=4.334^\circ$\\
\begin{tabular}{|c|c||c|c|c|c|c|c|c||c|c|c|c|}
\hline
$\phi$ &$x_4$&$\theta_a$&$\theta_b$&$\theta_c$&$\theta_W$&$x_1$&$x_2$&$x_5$&$2\theta_{CP}$
&$P(\nu_e)$&$P(\nu_\mu)$ &$P(\nu_\tau)$  \\
\hline
0  &1.5&  7.8&118.7&115.3& 31.5&0.43244&1.421&1.262&203.4&0.231 &0.377 &0.391  \\
\hline
60 &1.5& 20.2&178.5&266.2& 17.0&0.51667&1.037&0.844&124.0&0.238 &0.375 &0.387  \\
\hline
\end{tabular}
$\theta_V=124.334^\circ$\\
\begin{tabular}{|c|c||c|c|c|c|c|c|c||c|c|c|c|}
\hline
$\phi$ &$x_4$&$\theta_a$&$\theta_b$&$\theta_c$&$\theta_W$&$x_1$&$x_2$&$x_5$&$2\theta_{CP}$
&$P(\nu_e)$&$P(\nu_\mu)$ &$P(\nu_\tau)$  \\
\hline
0  &1.1& 25.5&281.5&276.6& 27.1&0.43244&1.399&1.643&138.4&0.235 &0.377 &0.388  \\
\hline
60 &1.1& 21.5&157.4&118.5& 19.8&0.51667&1.221&1.155&239.2&0.234 &0.375 &0.391  \\
\hline
120&1.1& 30.3&290.5&169.5& 22.2&0.61111&0.696&1.048& 38.4&0.255 &0.371 &0.373  \\
\hline
\end{tabular}
$\theta_V=244.334^\circ$\\
\begin{tabular}{|c|c||c|c|c|c|c|c|c||c|c|c|c|}
\hline
$\phi$ &$x_4$&$\theta_a$&$\theta_b$&$\theta_c$&$\theta_W$&$x_1$&$x_2$&$x_5$&$2\theta_{CP}$
&$P(\nu_e)$&$P(\nu_\mu)$ &$P(\nu_\tau)$  \\
\hline
0  &1.1& 18.9&282.6&101.6&153.2&0.43244&1.343&1.685&136.2&0.233 &0.377 &0.389 \\
\hline
60 &1.1& 26.8&177.4&294.9&165.8&0.51667&1.369&0.975&247.4&0.234 &0.375 &0.391 \\
\hline
120&1.1& 27.4&315.6&355.2&150.4&0.61111&0.892&0.887& 31.2&0.256 &0.371 &0.372 \\
\hline
\end{tabular}
\end{center}
\caption{The detection probabilities of neutrino flavors and argument of $K_{12}$.
Without loss of generality, we can define $(x_1,x_2,x_4,x_5)\geq 0$ and
$180^\circ\geq (\theta_a,\theta_W)\geq 0^\circ$. All angles are given in the unit of degree and
$x_i$ are given in the unit of $0.1 \sqrt{\mbox{eV}}$.}
\end{table}


\section{Conclusion}

In this paper we consider $S_4$ flavor-symmetric extra U(1) model
with taking account of high energy neutrino flux observed by IceCube
and obtain following results.
If we specify dark matter is the heaviest RHN, the mass scale PeV is understood as follow.  
The structure of CKM matrix requires $O(10^{-2})$ flavor symmetry breaking and
the symmetry should be non-abelian to suppress flavor changing processes
induced by sfermion exchange. 
Due to this symmetry, two lighter RHNs form $S_4$-doublet and have same mass.
As the $O(10^{-2})$ flavor symmetry breaking solves the mass degeneracy of 
$S_4$-doublet RHN by $O(10^{-4})$, the mass of RHN should be 100TeV
for a successful resonant leptogenesis.
The 3-body decay of heaviest RHN with 10PeV mass generates PeV energy neutrino 
through the following decay of lighter RHN.
Non-thermal production of the heaviest RHN gives right amount of dark matter.
The information about leptogenesis may be extracted from neutrino flux.

Although the flavor symmetry is broken,
as the footprint of it is left in the degenerated mass spectra of almost inert-Higgs
and G Higgs and their superpartners, 
the existence of flavor symmetry may be testable for LHC or future colliders.


\end{document}